\documentclass[
    ,final            % use final for the camera ready runs
  ]
  {aipproc}

\layoutstyle{6x9}

\begin{document}

\title{Depinning and wetting in nonequilibrium systems}

\author{F. de los Santos$^{*,\dagger}$, M.M. Telo da Gama$^\dagger$
and M.A. Mu\~noz$^{**}$}{ 
address={$^*$Center for Polymer Studies and Department of Physics, Boston
University, \\
590 Commonwealth Avenue, Boston, 02215 MA, USA \\
$^\dagger$ Centro de F\'\i sica Te\'orica e
Computacional da Universidade de Lisboa, \\
Avenida Professor Gama Pinto 2, P-1649-003 Lisboa, Portugal\\
$^{**}$Departamento de Electromagnetismo y F\'\i sica
de la Materia and Instituto Carlos I de F\'\i sica 
Te\'orica y Computacional, Universidad de Granada, \\ 18071 Granada, Spain}
}

\begin{abstract}
We present an extension of equilibrium wetting to nonequilibrium 
situations particularly suited to systems with anisotropic interactions. 
Both critical and complete wetting transitions were found
and characterized. We have identified a region in the space of parameters
(temperature and chemical potential) where the wet and non-wet
phases coexist. Emphasis is made on the analogies and differences
between equilibrium and nonequilibrium wetting.
\end{abstract}

\maketitle

%%%%%%%%%%%%%%%%%%%%%%%%%%%%%%%%%%%%%%%%%%%%
%% MAINMATTER
%%%%%%%%%%%%%%%%%%%%%%%%%%%%%%%%%%%%%%%%%%%%

\section{Equilibrium wetting}

Imagine that a small amount of liquid is poured on a substrate. 
At two-phase equilibrium,
i.e. a static situation where the liquid is at equilibrium with
its vapor, it may happen that the liquid does not coat 
the substrate, in which case it beads as droplets characterized by a
contact 
angle $\alpha$ as shown in figure 1. 
This is called partial wetting and the conctact angle is related to 
the surface tensions, $\sigma$, of the intervening intefaces 
through Young's formula (dating back to 1805) 
$\sigma_{sv} = \sigma_{sl}+ \sigma_{lv}\cos \alpha$, where
$\sigma_{sv}$ is the substrate/vapor surface tension and so on.
If, by contrast, the liquid spreads over the substrate and 
coats it uniformly (zero contact angle), 
the substrate is said to be wet by the liquid.
A wetting transition occurs when, by changing the temperature,
the substrate changes from a partially wet to wet state. 

It is instructive to study the same phenomenon from a 
different point of view. Figure 2 depicts the  
phase diagram of a pure substance. It is clear from 
the above discussion that if wetting is to occur, the system has to 
be at liquid/vapor coexistence, with an arbitrary fraction
in the liquid with the remainder in the vapor phase. The effect due to the
presence of a substrate 
that adsorbs preferentially the liquid is also displayed in Figure
2: 
above a certain temperature, $T_w$, called the wetting temperature,
the substrate is wet, while for $T<T_w$ the substrate is not wet at
coexistence.
This is illustrated on the right of the same figure where 
the thickness of the wetting layer, $h$, is displayed as a function of 
the temperature and the chemical potential difference between the 
liquid and vapor phases, $\mu$, along three different paths:
(1) the substrate remains not wet when coexistence is reached;
(2) the thickness of the wetting layer diverges continuously as
coexistence is approched from the gas phase (this is termed 
$complete \ wetting$); (3) as $T$ approches $T_w$ $at$ coexistence
the  thickness of the wetting layer may either diverge 
continuoulsy (denoted $critical \ wetting$) or discontinuously at
$T_w$ (denoted $first-order \ wetting$).

Assuming the system free energy to be a functional, ${\cal H}$, 
solely of the height $h({\bf x})$ of the liquid/vapor interface above
the substrate coordinate ${\bf x}$, then \cite{review} 

\begin{equation}
{\cal H}= \int dx \ \Bigg[ {\sigma \over 2} (\nabla h)^2
+V\big(h(x)\big) -\mu h \Bigg]
\end{equation}
where $\sigma$ is the surface tension and $V(h)$ accounts for 
the effective potential between the substrate and the interface. 
If all the interactions are short-ranged, then it can be proved 
that for large $h$ $V$ has the form $V(h)=b(T) e^{-h/\xi}+e^{-2h/\xi}$, 
with $b(T)$ proportional to $T_w$ and $\xi$ being the bulk 
correlation length \cite{review}. 

One obtains a dynamic model of wetting by relating the time derivative of 
$h$ with (minus) the functional deivative of $\cal H$ through,

\begin{equation}
{\partial h({\bf x},t) \over \partial t}=-{\delta {\cal H}\over \delta h}
+\eta({\bf x},t) =
\sigma \nabla^2 h
-{\partial V(h) \over \partial h} +\mu +\eta,
\label{equilibrium}
\end{equation}
where $\eta$ is a Gaussian withe noise with mean 
$\langle \eta ({\bf x},t) \rangle =0$
and correlations
$\langle \eta ({\bf x},t) \eta ({\bf x}',t') \rangle
= 2D\delta({\bf x}-{\bf x}')\delta(t-t')$.
In this context wetting appears in the guise of an unbinding 
transition and its phenomenology is described by the long-time
behavior of the solutions of \eqref{equilibrium}
as follows: (i) liquid/vapor coexistence obtains at  
$\mu= \mu_c=0$, since for this value of the chemical potential
difference the free interface does not move on average irrespective of its 
initial position. This is no longer the case when the system 
is at contact
with a substrate. Under these circunstances there is a value $b_w$,
(proportional to) the temperature, above which
$\langle h \rangle \to \infty$ as $t \to \infty$ at coexistence, $\mu=0$
(critical wetting). Complete wetting corresponds to the divergence of
$\langle h \rangle$ as $t \to \infty$ and $\mu \to 0^-$
for values of $b>b_w$\cite{lipowsky}.

\begin{figure}
\includegraphics[height=.15\textheight]{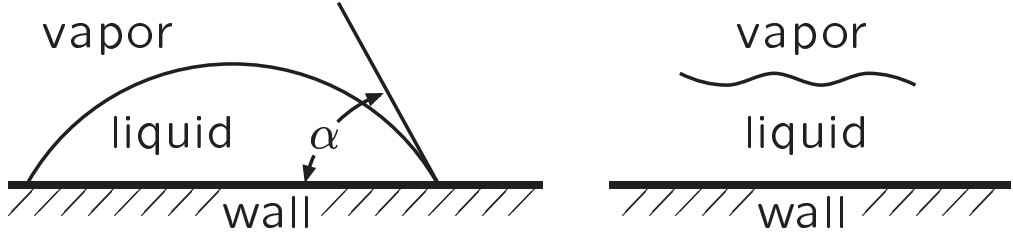}
\caption{Left, liquid droplet at equilibrium with its vapor on a 
(non-wet) planar substrate (wall); right, same situation for a uniform
coating (wetting) of the substrate.}
\end{figure}

\begin{figure}
\includegraphics[height=.35\textheight]{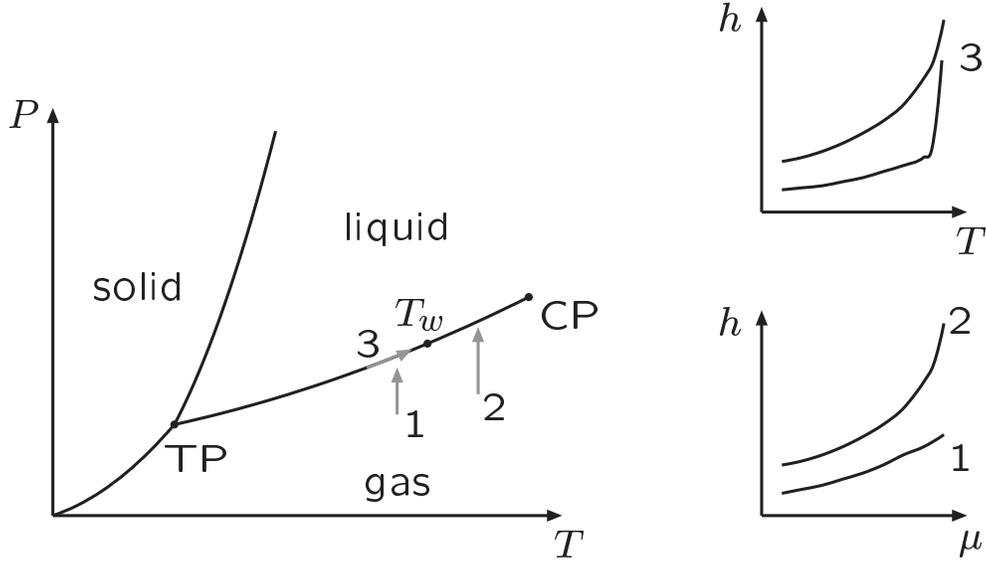}
\caption{Left, pressure vs. temperature phase diagram of a pure substance.
$TP$, $CP$ and $T_w$ stand for the triple point, critical point and
wetting temperature, respectively;
right, thickness of the wetting layer as a function of the temperature
and chemical potential difference for the paths indicated on the left.}
\end{figure}

\section{Nonequilibrium wetting}

Consider the following Langevin equation, 

\begin{equation}
{\partial h ({\bf x},t)\over \partial t} =\sigma
\nabla^2 h + \lambda \big(\nabla h \big)^2
-{\partial V(h) \over \partial h} +\eta
\label{noneq}
\end{equation}
which differs from \eqref{equilibrium} by the presence a 
new non-linearity, namely, 
the {\em KPZ} term $\lambda (\nabla h)^2$. 
In this section we will study the 
wetting properties of a system described by \eqref{noneq}.
The absence of an equilibrium Hamiltonian for \eqref{noneq} 
shows that it is a genuine non-equilibrium equation and justifies the 
title of 
this section.

There are a number of good reasons to include the KPZ term.
It may be intrepreted as a force acting on the tilted parts of the interface
in the direction of growth and, therefore, it is relevant 
in systems with anisotropic interactions, where the growth of tilted
interfaces may depend on their orientation. In fact, it governs the 
growth of crystals from atomic beams when desorption is allowed
\cite{villain}. Further, a renormalization group study 
has shown that it is always generated, except when excluded by symmetry, 
when elastic objects depin in the presence of anisotropy \cite{wiese}. 
Finally, lattice models of nonequilibrium wetting 
seem to be controlled by the KPZ non-linearity \cite{hinrichsen}.

Investigating the wetting behavior of \eqref{noneq} requires carrying out
the steps outlined in the previous section, 
a programme we have completed for $\lambda = -\sigma=-1$. 
We would like to stress that while the value of the surface tension is 
irrelevant, the sign of $\lambda$ determines the behavior of
\eqref{noneq}.
By contrast with the equilibrium system, bulk coexistence no longer
obtains at $\mu=0$. Rather, $\mu_c = \langle (\nabla h)^2 \rangle$
that for one-dimensional substrates  
is given approximately by $\mu_c = -D\lambda /(2\sigma \Lambda)$,
where $\Lambda$ is a lattice cutoff \cite{krug}. In higher 
dimensionalities one has to resort to numerical methods to obtain $\mu_c$.

Our findings are sketched in figure 3. Nonequilibrium critical wetting
occurs as $b \to b_w = -0.32 \pm 0.05$ along path 1. 
$\langle h \rangle$ diverges as $|b-b_w|^\beta$, with 
$\beta= -2.6 \pm 0.2 $.
This value differs from that of equilibrium wetting, $\beta=-1$ \cite{review}
thereby defining a new universality class as expected. Along path 2
nonequilibrium complete wetting is observed as $\mu \to \mu_c$ with
$b>b_w$. The associated exponent in this case is $- 0.41$, with
error bars that exclude the equilibrium value $\beta = -1/3$.

As we have said before, no wetting transition can occur below the wetting 
temperature. Pushing a little bit further (within this interfacial
model) one would expect a depinning transition when crossing the boundary 
$\mu=\mu_c$, for
$\mu>\mu_c$ is the realm of the liquid phase. Interestingly enough,
it turns out that the vapor phase is stable up to
$\mu=\mu^*(b)>\mu_c$ (path 3 of figure 3). In fact,
within the region delimited by the dashed lines in figure 3, 
the dipinned and pinned phases coexist.
As in equilibrium, this means that the system will either exhibit 
a vapor (pinned) or liquid (depinned) phase depending on the initial 
conditions. The fact that the coexistence region is finite
rather than a line is, however, a nonequilibrium effect. 
The microsocopic mechanism underlying this behavior 
is illustrated on the left of figure 3: when a bound interface
makes a large excursion away from the substrate, marking (in
principle) the onset of the depinning transition,
the fluctuation acquires a triangular shape (pyramidal for
two-dimensional substrates), that is pushed down due to the KPZ
term and, eventually, suppressed. This mechanism operates 
in a finite region of the space of parameters $(T,\mu)$
and explains why the coexistence region is finite 
\cite{hinrichsen,nos,marsili}.

\begin{figure}
\includegraphics[height=.25\textheight]{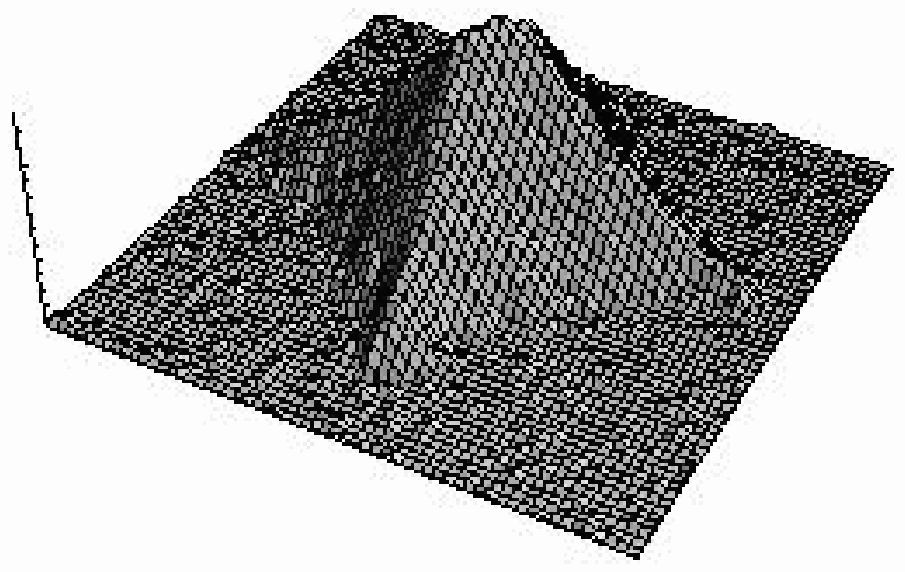}
\includegraphics[height=.25\textheight]{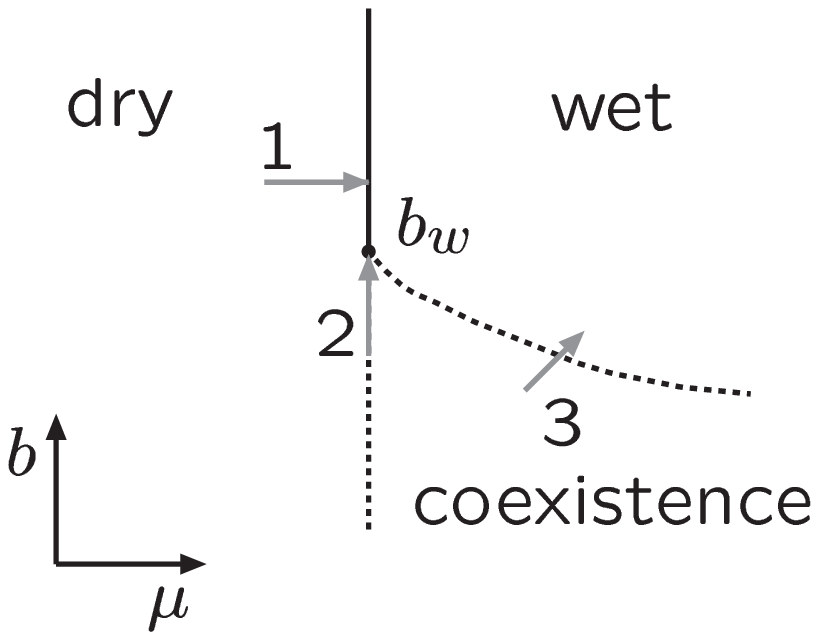}
\caption{Right, phase diagram from eq. \eqref{noneq}; left,
typical structures observed in the coexistence region for a
two-dimensional
sustrate.}
\end{figure}

%%%%%%%%%%%%%%%%%%%%%%%%%%%%%%%%%%%%%%%%%%%%

\section{Conclusions}
We have proposed and solved a continuum model for nonequilibrium wetting. 
It consists of a dynamic version of a well known 
equilibrium wetting Hamiltonian supplemented with a KPZ non-linear term.
We focussed on short-ranged forces and negative
non-linearities. We found and characterized in detail the nonequilibrium 
counterparts of critical and complete wetting transitions. A finite region
of coexistence of wet and nonwet phases, the existence of triangular
(pyramidal) patterns within this region and the relation to depinning
transitions below the wetting temperature have also been discussed.

%%%%%%%%%%%%%%%%%%%%%%%%%%%%%%%%%%%%%%%%%%%%%%%%
%% BACKMATTER
%%%%%%%%%%%%%%%%%%%%%%%%%%%%%%%%%%%%%%%%%%%%%%%%

\begin{theacknowledgments}
We acknowledge financial support from the E.U. through Contract No.
ERBFMRXCT980183, by the Ministerio de Ciencia y
Tecnolog\'\i a (FEDER) under project BFM2001-2841 and from the
Funda\c c\~ao para a Ci\^encia e a Tecnologia, contract SFRH/BPD/5654/2001.
\end{theacknowledgments}

%%%%%%%%%%%%%%%%%%%%%%%%%%%%%%%%%%%%%%%%%%%%%%%%
%% You may have to change the BibTeX style below, depending on your
%% setup or preferences.
%%
%% If the bibliography is produced without BibTeX comment out the
%% following lines and see the aipguide.pdf for further information.
%%
%% For The AIP proceedings layouts use either
%%%%%%%%%%%%%%%%%%%%%%%%%%%%%%%%%%%%%%%%%%%%

%\bibliographystyle{aipproc}   % if natbib is available
%\bibliographystyle{aipprocl} % if natbib is missing

%%%%%%%%%%%%%%%%%%%%%%%%%%%%%%%%%%%%%%%%%%%
%% You probably want to use your own bibtex database here
%%%%%%%%%%%%%%%%%%%%%%%%%%%%%%%%%%%%%%%%%%%
%\bibliography{sample}
%\begin{references}

%%%%%%%%%%%%%%%%%%%%%%%%%%%%%%%%%%%%%%%%%%%
%% Just a reminder that you may have to run bibtex
%% All of it up to \end{document} can be removed
%% if you don't like the warning.
%%%%%%%%%%%%%%%%%%%%%%%%%%%%%%%%%%%%%%%%%%%
\IfFileExists{\jobname.bbl}{}
 {\typeout{}
  \typeout{******************************************}
  \typeout{** Please run "bibtex \jobname" to optain}
  \typeout{** the bibliography and then re-run LaTeX}
  \typeout{** twice to fix the references!}
  \typeout{******************************************}
  \typeout{}
 }

\end{document}